\documentclass[final,3p,times,twocolumn]{elsarticle}
\usepackage{graphicx}
\usepackage{amsmath, amsthm, amssymb}
\usepackage{epsfig}

\newcommand{\beq}{\begin{equation}}
\newcommand{\beqa}{\begin{eqnarray}}
\newcommand{\eeq}{\end{equation}}
\newcommand{\eeqa}{\end{eqnarray}}
\newcommand{\abs}[1]{\left\vert#1\right\vert}

\renewcommand{\d}{{\rm d}}
\newcommand{\e}{{\rm e}}

\renewcommand{\max}{{\rm max}}
\newcommand{\mean}[1]{\langle#1\rangle}
\newcommand{\petit}{{\hskip -7pt}}
\newcommand{\avant}{{\hskip 7pt}}
\newcommand{\prob}{\mathop{\rm Prob}\nolimits}
\newcommand{\sat}{{\rm sat}}
\renewcommand{\th}{{\rm th}}
\newcommand{\w}{\hat}
\newcommand{\var}{\mathop{\rm var}\nolimits}
\renewcommand{\P}{\prob}
\newcommand{\R}{{\cal R}}

\journal{Journal of Theoretical Biology}

\begin{document}

\begin{frontmatter}

\title{Fluctuation effects in metapopulation models:\\
percolation and pandemic threshold}

\author{Marc Barth\'elemy}
\author{Claude Godr\`eche}
\author{Jean-Marc Luck}

\address{Institut de Physique Th\'eorique\\
CEA Saclay, and URA 2306, CNRS, 91191 Gif-sur-Yvette, France}

\begin{abstract}
Metapopulation models provide the theoretical framework for
describing disease spread between different populations connected by
a network. In particular, these models are at the basis of most
simulations of pandemic spread. They are usually studied at the
mean-field level by neglecting fluctuations. Here we include
fluctuations in the models by adopting fully stochastic descriptions
of the corresponding processes. This level of description allows to
address analytically, in the SIS and SIR cases, problems such as the
existence and the calculation of an effective threshold for the
spread of a disease at a global level. We show that the possibility
of the spread at the global level is described in terms of (bond)
percolation on the network. This mapping enables us to give an
estimate (lower bound) for the pandemic threshold in the SIR case
for all values of the model parameters and for all possible
networks.

\end{abstract}


\end{frontmatter}

\section{Introduction}

Modeling the spread of a disease between different populations
connected by a transportation network is nowadays of crucial
importance. Global spread of pandemic influenza is an illustrative
example of the importance of this problem for our modern societies. In
the context of pandemic spread and airline transportation networks,
Rvachev and Longini~\cite{Rvachev:1985} proposed a model, now called a
metapopulation model, which was used to describe the spread of
pandemic influenza~\cite{Longini:1988,Grais:2003,Grais:2004,Colizza:2006a},
SARS~\cite{Hufnagel:2004,Colizza:2007}, HIV~\cite{Flahault:1991}, and very
recently swine flu~\cite{Vespi:2009a,Vespi:2009b}. This model
describes homogeneously mixed populations connected by a
transportation network. Despite its successes, very few theoretical
studies are available. Numerical studies~\cite{Colizza:2006b} showed
that weight heterogeneity played a major role in the predictability by
creating epidemic pathways which compensate the various scenario
possibilities due to hubs~\cite{Colizza:2007}.

A major advance in our understanding of metapopulation models was done
in the series of papers by Colizza, Pastor-Satorras and
Vespignani~\cite{Colizza:2007a,Colizza:2007b,Colizza:2008}. In
particular, these authors studied the condition at which a disease
which can spread in an isolated population (and has therefore a basic
reproductive rate $\R_0>1$ (see below)) can invade the whole
network. In this respect, these authors defined an effective
reproductive rate $\R_*$ at the global level and showed that its
expression for a random network, in the SIR case with $\R_0\rightarrow
1$, and with a travel rate $p$, is under a mean-field
approximation~\cite{Colizza:2007a}
\begin{equation}
\R_*=(\R_0-1)\frac{\langle k^2\rangle -\langle k\rangle}{\langle k\rangle^2}
\frac{pN\alpha}{\mu},
\label{eq:rstar}
\end{equation}
where the brackets $\langle\cdot\rangle$ denote the average over the
configurations of the network, while $\alpha$ represents the fraction
of infected individuals in a given city during the epidemic, $1/\mu$
is the mean time an individual stays infected, and $N$ is the mean
population of a city. For $\R_* > 1$ the disease spreads over an
extensive number of cities. For scale-free networks, the second moment
$\mean{k^2}$ being very large, travel restrictions (i.e., decreasing
$p$) have a very mild effect. Equation~(\ref{eq:rstar}) relies on a
number of assumptions: the network is uncorrelated and tree-like, $\R_0$
is close to $1$, and more importantly the number of infected
individuals going from one city to another is assumed to be given by
the constant $d=pN\alpha/\mu$. This last expression however neglects
different temporal effects. Indeed, in the SIR case we can distinguish
two different time regimes: the first regime of exponential growth
corresponding to the outbreak, and the recovery regime where, after
having reached a maximum, the number of infected individuals falls off
exponentially. The number of infected individuals going from one city
to another thus depends on different factors and can be very different
from the expression for $d$ given above.

In the present work, we consider the stochastic version of the
metapopulation model. We investigate in detail the number of infected
individuals going from one site to the other. This allows us to
address the problem of the condition for propagation at a global scale
without resorting to mean-field approximations or $\R_0\rightarrow 1$.
We will show that the latter can be recast in terms of a bond
percolation problem.

So far most of the simulations on the metapopulation model were done
in essentially two ways. The first approach consists in using
Langevin-type equations, with a noise term whose amplitude is
proportional to the square root of the
reaction term~\cite{Gardiner:2004,Colizza:2006a}. In this approach, the evolution of
the various populations (infected, susceptibles, removed) is described
by a set of finite difference equations, with noise and with traveling
random terms describing the random jumps between
cities~\cite{Colizza:2006a}. This level of description suffers from
several technical difficulties when the finite-difference equations
are iterated: truncation of noise, non-integer number of individuals,
etc., which render its numerical implementation difficult. Another
possibility consists in adopting an agent-based approach, i.e.,
simulating the motion of each individual~\cite{Colizza:2006a}. This
simulation is free of ambiguities but requires a lot of memory and CPU
time.

In contrast, in the present work we will put forward an intermediate
level of description, which is free from the technical difficulties of
the Langevin-type approach (it does not require to iterate
finite-difference equations), and which does not require large amounts
of CPU. This approach consists in describing the process at the
population level, considering that individuals are
indistinguishable. In Statistical Mechanics, this level of description
is used for urn models or migration processes
(see~\cite{Godreche:2002,Godreche:2007} for reviews). Let us
take for definiteness the example of the stochastic SIS epidemic
process. For a single isolated city with population $N$, this process
(see for example~\cite{Allen:2003}) is described in terms of two
variables: $S(t)$ (number of susceptible individuals) and $I(t)$
(number of infected individuals), which evolve in continuous time as
\beq\label{sis1} (I,S)\rightarrow\begin{cases}
(I+1,S-1)& \text{with rate}\ \ \lambda SI/N,\\
(I-1,S+1)& \text{with rate}\ \ \mu I,
\end{cases}
\eeq
where $S(t)+I(t)=N$ is conserved, and with initial condition $I(t=0)=I_0$.
The first reaction corresponds to the meeting of a susceptible individual with
an infected individual, resulting in two infected ones. The second reaction
corresponds to the recovery of an infected individual into a susceptible one.
The SIR case has an analogous definition (see Section~3).

For two or more cities, we have to take into account the
process of traveling between different cities. This diffusion process is described
at the microscopic level by the jumping of individuals from one city to
another. Let $i$ and $j$ be two neighboring cities.
As described above, at
the level of populations (i.e., such that the individuals are indistinguishable),
the process inside each city is given by~(\ref{sis1}), while the elementary
traveling processes between these two cities are:
\beq\label{eq:travel}
\begin{matrix}
&\hfill(I_i,I_j)\rightarrow(I_i-1,I_j+1)\hfill& \text{with rate}\ \ p_{ij}I_i,\hfill\\
&\hfill(S_i,S_j)\rightarrow(S_i-1,S_j+1)\hfill& \text{with rate}\ \ p_{ij}S_i,\hfill
\end{matrix}
\eeq
where $p_{ij}$ is the probability per unit time and per individual to jump from city $i$
to city $j$.

As said above, the numerical implementation of these processes is easy,
and it allows us to make simulations with large numbers of large cities,
in contrast with agent-based simulations.
In the following we will restrict ourselves to uniform symmetric diffusion,
defined by the constant rates $p_{ij}=p$ if cities $i$ and $j$ are neighbors.
These rules are compatible with a stationary state
where all cities have the same population ($I_i+S_i=N_i=N$) on average.
We will consider the initial condition $I_i(t=0)=I_0\delta_{i0}$,
where $I_0$ is the initial number of infected individuals,
which all live in city~0 at time $t=0$.
All the numerical simulations of the SIS and SIR cases will be performed
with $\lambda=0.3$ and $\mu=0.1$, so that $\R_0=3$ (see~(\ref{rdef})).

The outline of the paper is as follows.
In Section~2 we consider the SIS case in different geometries, starting
with the case of a single city, then proceeding to the case of two cities,
and finally discussing the propagation on an infinite one-dimensional array of cities.
As the number of infected individuals in a given city
saturates to a non-zero fraction of the total population,
the metapopulation model will always experience a
pandemic spread and ballistic propagation. It is however instructive to study this
model for the following reasons: (i) it displays the same phase of exponential growth
as the SIR model, but with the great simplification that
only two compartments ($S$ and $I$) are present, allowing
analytical approaches; (ii) the SIS model has connections with the
FKPP equation~\cite{Fisher:1937,Kolmo:1937},
which is still currently the subject of intense activity.
In Section~3 we consider the SIR case along the same lines.
We successively consider the simple cases of one and two cities,
before addressing the question of the propagation on various extended structures
(one-dimensional chain, square lattice, Cayley tree, uncorrelated scale-free network).
The key difference with the SIS case is that the time integral
of the number of infected individuals in a given city is now finite.
We demonstrate that the latter property
implies the existence of an effective pandemic threshold
for the spread of the disease over the whole network.
We also obtain an estimate (lower bound) for this pandemic threshold
as a function of all the model parameters
and of the bond percolation threshold on the underlying geometrical structure.
For scale-free networks our prediction is virtually identical
to that of~\cite{Colizza:2007a}.

\section{Metapopulation model in the SIS case}

In this section we discuss the metapopulation model in its
stochastic form, on the example of the SIS process.
We start with the case of a single city, then proceed to the case of two cities,
which will be the building block of our analysis of the propagation of the epidemic
on extended structures.
We finally discuss the ballistic propagation on a one-dimensional array of cities.

\subsection{Stochastic SIS model for one city}

We first present a discussion of the stochastic version of the SIS model for one city.

Let $N$ be the population of the city and $I(t)$ the number of infected
individuals at time $t$.
As can be seen from the defining reactions~(\ref{sis1}), the SIS process can be
described by the single random variable $I(t)$, which experiences a biased
continuous-time random walk on the interval $(0,N)$,
with variable rates which only depend on the
instantaneous position $I(t)=k$ of the walker.
This process is known as a birth-and-death process
in the probabilistic literature~\cite{feller}.
Let us denote by $\lambda_k$ and $\mu_k$ the jump rates of the walker,
respectively to the right ($k\to k+1$) and to the left ($k\to k-1$).
Here we have
\beq\label{lam}
\lambda_k=\lambda k\left (1-\frac{k}{N}\right),\quad\mu_k=\mu k.
\eeq
Hence the origin ($k=0$) is absorbing while the right end ($k=N$) is reflecting.
The walker moves in an effective confining potential, with a restoring force
bringing it back to the quasi-equilibrium position corresponding to the
equality of the rates $\lambda_k$ and $\mu_k$.
This position can be interpreted as the number of infected individuals at saturation.
It reads
\beq
I_\sat=N\left(1-\frac{1}{\R_0}\right).
\label{isatdef}
\eeq
The so-called basic reproductive number is defined as
\beq
\R_0=\frac{\lambda}{\mu},
\label{rdef}
\eeq
and we will hereafter focus on the case of interest where $\R_0$ is larger than 1, which is the condition for the occurrence of an outbreak at the level of a single city.
The quasi-equilibrium state around $I_\sat$ is however metastable,
since the walker is deemed
to be ultimately absorbed at the origin, which corresponds to the extinction of
the epidemic in the city. The characteristic absorption time
is however exponentially increasing with the population $N$ of the city,
as will be shown below.

\begin{figure}[ht!]
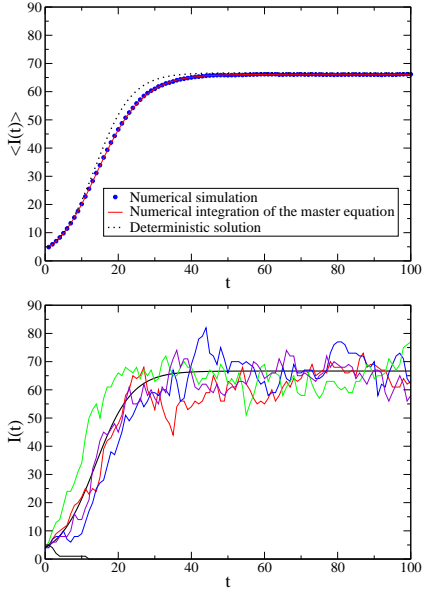

\centering
\begin{tabular}{c}
\epsfig{file=figure1b.eps,width=0.7\linewidth,clip=} \\
\epsfig{file=figure1a.eps,width=0.7\linewidth,clip=}
\end{tabular}
\caption{
(Color online).
Top: Mean number of infected individuals in the stochastic SIS
process for one city, conditioned on no extinction of the epidemic.
Smooth red line: numerical integration of the master equation~(\ref{eq:master}).
Blue circles: Results of a numerical simulation.
Dotted black line: deterministic solution~(\ref{eq:determ}).
Bottom: Five different realizations of the stochastic SIS process.
Smooth black line: deterministic solution~(\ref{eq:determ}).
($\R_0=3$, $N=100$, $I_0=4$.)}
\label{fig1}
\end{figure}

Let us denote by $f_k(t)=\prob(I(t)=k)$ the probability for the number of
infected individuals to be equal to the integer $k$ at time $t$.
This probability contains all the information on the one-time properties of the
process (see~\cite{Godreche:2002,Godreche:2007} for reviews).
Its temporal evolution is given by the following master equations,
characteristic
of birth-and-death processes~\cite{karlin,kendall}:
\begin{eqnarray}\label{eq:master}
\avant\frac{\d f_k}{\d t}\petit
&=&\petit\lambda_{k-1}f_{k-1}+\mu_{k+1}f_{k+1}-(\lambda_k+\mu_k)f_k,\nonumber\\
\frac{\d f_0}{\d t}\petit&=&\petit\mu_1 f_1,\nonumber\\
\frac{\d f_N}{\d t}\petit&=&\petit\lambda_{N-1}f_{N-1}-\mu_N f_N.
\end{eqnarray}
The nonlinear dependence of the rates on the position $k$
renders their analytical study difficult.
For example the evolution equation of the first moment reads
\beq\label{eq:momt}
\frac{\d \langle I\rangle}{\d t}=(\lambda-\mu)\langle I\rangle-
\frac{\lambda}{N}\langle I^2\rangle,
\eeq
whereas the equation for $\langle I^2\rangle$ itself involves $\langle I^3\rangle$
and so on.
However, eqs.~(\ref{eq:master}) are easily implemented numerically,
for a given initial condition.
In Figure \ref{fig1} (top) we show
a plot of $\langle I\rangle$ obtained by a numerical integration of
eqs.~(\ref{eq:master}).
We also plot the same quantity as obtained from a numerical simulation of the
SIS process.
The data were obtained for $\lambda=0.3$, $\mu=0.1$, $N=100$, and $I_0=4$,
discarding histories leading to absorption.
The two sets of data are indistinguishable one from the other.
Finally we also show for comparison the deterministic (or mean-field) expression
of $I(t)$ (see~(\ref{eq:determ}) below).
Figure \ref{fig1} (bottom) shows five different realizations of the stochastic process for a single city, together with the deterministic solution~(\ref{eq:determ}).

The deterministic (or mean-field) approximation consists in neglecting any correlations.
For example, in~(\ref{eq:momt}), it amounts to replacing $\mean{I^2}$ by $\mean{I}^2$.
This approach is a priori legitimate if the population $N$ of the city is large,
so that the number of infected individuals $I(t)$ is also typically large
enough to be considered as a deterministic observable, instead of a random variable.
This approximation thus reduces the SIS model for one city
to a deterministic dynamical system.
Then~(\ref{eq:momt}) becomes the following dynamical equation
for the temporal evolution of $I(t)$:
\beq\label{eq:sismf}
\frac{\d I}{\d t}=(\lambda-\mu)I-\frac{\lambda}{N}I^2,
\eeq
the solution of which can be cast into the form
\beq\label{eq:determ}
\frac{1}{I(t)}=\frac{\e^{-(\lambda-\mu)t}}{I_0}+\frac{1-\e^{-(\lambda-\mu)t}}{I_\sat}.
\eeq

As demonstrated by Figure~\ref{fig1}, the model exhibits an exponential growth regime
followed by a saturation regime.
The deterministic approximation gives an accurate global description of $I(t)$,
even for rather small populations.
It however misses the effect of fluctuations in the process,
that we now study in both regimes successively.

\subsubsection*{Exponential growth regime}

The regime of exponential growth formally corresponds to taking the infinite $N$ limit
at fixed time $t$.
In this limit, (\ref{eq:momt}) and (\ref{eq:sismf}) simplify, and their common solution
reads
\beq
\mean{I(t)}=I_0\,\e^{(\lambda-\mu)t}.
\label{imean}
\eeq
Furthermore, the rates become linear in $k$, i.e., $\lambda_k=\lambda k$ and $\mu_k=\mu k$,
reducing the model to a solvable birth-and-death process~\cite{feller,kendall}.
The master equations for the~$f_k$ read
\begin{eqnarray}
\nonumber
\avant \frac{\d f_k}{\d t}\petit&=&\petit\lambda (k-1)f_{k-1}+\mu
(k+1)f_{k+1}-(\lambda+\mu)kf_k,\\
\frac{\d f_0}{\d t}\petit&=&\petit\mu f_1.
\end{eqnarray}
The generating function
\beq
G(s,t)=\mean{s^{I(t)}}=\sum_{k\geq 0}s^kf_k(t)
\eeq
satisfies
\begin{equation}
\frac{\partial G}{\partial t}=(s-1)(\lambda s-\mu)\frac{\partial G}{\partial s},
\end{equation}
the solution of which is obtained with the method of characteristics and is
\begin{equation}\label{gst}
G(s,t)= \left(\frac {\mu (s-1)-(\lambda s-\mu)\e^{-(\lambda-\mu)t}}
{\lambda (s-1)-(\lambda s-\mu)\e^{-(\lambda-\mu)t}} \right)^{I_0}.
\end{equation}
In particular, the probability for the number of infected individuals to be zero
at time $t$, $f_0(t)=G(0,t)$, reads~\cite{kendall}
\begin{equation}
f_0(t)=\left(
\frac{\mu (1-\e^{-(\lambda-\mu)t})}
{\lambda-\mu\,\e^{-(\lambda-\mu)t}}
\right)^{I_0}.
\end{equation}
For $t\to\infty$ we recover the well-known extinction probability~\cite{Bailey}
\begin{equation}
p_{\rm extinct}=f_0(\infty)=\frac{1}{\R_0^{I_0}}.
\label{pextinct}
\end{equation}
If $\R_0<1$ this extinction probability is equal to 1.
Hence the condition $\R_0>1$ ensures the possibility of a local outbreak at the level of a single city.
The expressions of the moments of $I(t)$ can be extracted from~(\ref{gst}).
We thus recover~(\ref{imean}), while the variance reads
\begin{equation}\label{var_deter}
\var I(t)=\frac{\R_0+1}{\R_0-1}\,\frac{\mean{I(t)}}{I_0}\,(\mean{I(t)}-I_0).
\end{equation}
In the late stages of the exponential growth regime,
the relative variance is proportional to $1/I_0$:
\beq
\frac{\var I}{\mean{I}^2}=\frac{\R_0+1}{\R_0-1}\,\frac{1}{I_0}.
\label{vargrow}
\eeq

\subsubsection*{Saturated regime}

After the phase of exponential growth is over,
the mean number of infected individuals saturates at the value $\mean{I(t)}=I_\sat$
given by ~(\ref{isatdef}), as seen in Figure~\ref{fig1}.
The time to reach saturation therefore scales as $t_\sat\approx\ln(N/I_0)/(\lambda-\mu)$.

The saturated state is actually a metastable state, rather than a genuine stationary state.
For finite $N$, the system will indeed always end in its absorbing state at $k=0$.
The distribution $f_k$ of the fluctuating number of infected individuals
in the metastable state can be evaluated by cutting the link from $k=1$ to $k=0$
which is responsible for absorption.
The process thus modified reaches an equilibrium state at long times.
In the latter state the distribution $f_k$ satisfies detailed balance, that is
\beq
\mu_{k+1}f_{k+1}=\lambda_k f_k,
\eeq
where $\lambda_k$ and $\mu_k$ are defined in~(\ref{lam}).
This equation yields
\beq
\frac{f_k}{f_1}
=\frac{1}{k\R_0}\frac{(N)!}{(N-k)!}\left(\frac{\R_0}{N}\right)^{k}\\
\sim\e^{N\,g(k/N)},
\label{gdef}
\eeq
where the large deviation function $g(x=k/N)$ reads
\beq
g(x)=x(\ln\R_0-1)-(1-x)\ln(1-x).
\eeq
As expected, the distribution $f_k$ is peaked around $k=Nx_c=I_\sat$ for $N$ large.
Indeed the function $g(x)$ takes its maximal value,
\beq
g_\max=\ln\R_0+\frac{1}{\R_0}-1,
\eeq
for $x=x_c=1-1/\R_0$.
We therefore have the exponential estimate $f_{k_c}/f_1\sim\e^{Ng_\max}$.
The lifetime of the metastable state can then be estimated,
in the spirit of the Arrhenius law,
as $\tau\sim1/f_1$.
It is therefore predicted to grow exponentially with the population as
\begin{equation}
\tau_N\sim\e^{Ng_\max}.
\end{equation}
For $\R_0=3$ we have $g_\max=0.431945$.

\begin{figure}[ht!]
\centering
\begin{tabular}{c}
\epsfig{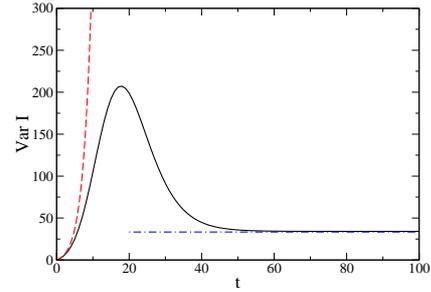}
\end{tabular}
\caption{
(Color online).
Variance of the number of infected individuals in the stochastic SIS
process for one city ($\R_0=3$, $N=100$, $I_0=4$).
Continuous line: numerical integration of the master equations (\ref{eq:master}).
Red dashed line: prediction (\ref{var_deter}).
Blue dot-dashed line: stationary value (\ref{vars}).}
\label{fig2}
\end{figure}

The expression~(\ref{gdef}) also yields an estimate for the Gaussian fluctuations
of $I(t)$ around its value $I_\sat$ in the saturated state.
Indeed, expanding $g(x)$ to second order around $x_c$, we obtain the estimate
\beq
\var I=\frac{N}{\R_0}
\label{vars}
\eeq
for the variance of the distribution of $I$ in the quasi-stationary state,
so that the reduced variance
\beq
\frac{\var I}{\mean{I}^2}=\frac{\R_0}{(\R_0-1)^2}\,\frac{1}{N}
\label{varstat}
\eeq
is proportional to $1/N$, whereas it was proportional to $1/I_0$ in the growth phase
(see (\ref{vargrow})).
These two results provide a quantitative confirmation that relative fluctuations
around the deterministic theory become negligible in all regimes,
as soon as the number of infected individuals is large.
Figure~\ref{fig2} shows a plot of $\var I$,
obtained by integration of the master equations (\ref{eq:master}).
The data for small and large times
are found to be in very good agreement with the estimates
(\ref{var_deter}) and (\ref{vars}), respectively.

\subsection{Including travel: two cities}

Consider now two cities, numbered 0 and 1.
The traveling process is described by~(\ref{eq:travel}).
We assume symmetric diffusion: $p_{01}=p_{10}=p$,
hence the system reaches a stationary state such that $N_0=N_1=N$.
We choose the initial condition $I_i(t=0)=I_0\delta_{i0}$.
We are primarily interested in the time of occurrence
of the outbreak of the epidemic in city~1.
This event is governed by the arrival of infected individuals
traveling from city~0 to city~1.
Returns of infected individuals from city~1 to city~0
can be neglected throughout.

We start by analyzing the distribution of the first arrival time $t_1$ of an
infected individual in city~1.
Let us denote by $Q(a,b)$ the probability that no infected individual exits
from city~0 in the time interval $(a,b)$, with in particular
$Q(0,t)=\prob(t_1>t)$.
We have
\beqa
\avant Q(0,t+\d t)\petit&=&\petit Q(0,t)Q(t,t+\d t)\nonumber\\
\petit&=&\petit Q(0,t)(1-pI_0(t)\d t),
\eeqa
where the expression in the last parentheses is the probability that no
infected individual exits from city~0 in the infinitesimal time interval $(t,t+\d t)$.
We thus obtain
\beq
\frac{\d Q(0,t)}{\d t}=-p I_0(t) Q(0,t),
\eeq
and so
\beq
\prob(t_1>t)=\e^{-\Lambda(t)},
\label{eq:arrivaltime}
\eeq
with
\beq
\Lambda(t)=p\int_0^t \d\tau\,I_0(\tau).
\label{lambdadef}
\eeq An alternate derivation of this result
Eqs. (\ref{eq:arrivaltime},\ref{lambdadef}) is given
in~\cite{Gautreau:2008}.  We thus conclude that the number $N(t)$ of
arrivals of infected individuals in city~1 in the time interval
$(0,t)$ is a Poisson process, for which the rate is itself a
stochastic process, equal to $\d\Lambda/\d t=p I_0(t)$, as the
integrated rate is $\Lambda(t)$.  Therefore \beq
\prob(N(t)=n)=\e^{-\Lambda(t)}\,\frac{\Lambda(t)^n}{n!}\quad(n=0,1,\ldots).
\eeq This generalization of the Poisson process is known in the
literature as the Cox process~\cite{Cox}.  Hereafter, in order to
simplify the presentation, we will implicitly assume that
probabilities are conditioned on a single stochastic history $I_0(t)$.

\begin{figure}[ht!]
\centering
\begin{tabular}{c}
\epsfig{file=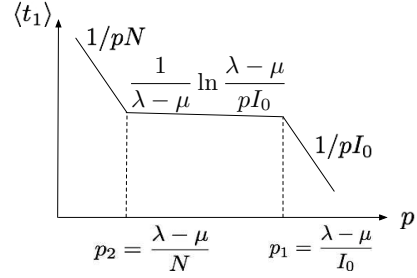,width=0.7\linewidth,clip=}
\end{tabular}
\caption{
Schematic (log-log) representation of the three different
regimes for the mean arrival time $\mean{t_1}$ as a function of $p$.}
\label{fig3}
\end{figure}

The mean first arrival time can easily be derived from~(\ref{eq:arrivaltime}).
It has the simple form
\begin{equation}\label{t1_cox}
\langle t_1\rangle=\int_0^\infty \d t\,\e^{-\Lambda(t)}.
\end{equation}
This exact expression can not be written in closed form,
even if $I_0(t)$ is given its deterministic value~(\ref{eq:determ}).
The scaling properties of $\langle t_1\rangle$ at large $N$
can however be obtained using the following simple arguments.
In the large $p$ regime, we expect $\langle t_1\rangle \approx 1/(pI_0)$,
as the number of infected individuals has hardly changed from its initial value $I_0$
in time $\langle t_1\rangle$.
In the opposite regime, where~$p$ is very small,
we have $I_0(t)\approx I_\sat$ (see~(\ref{isatdef})),
and thus $\langle t_1\rangle\approx 1/(pI_\sat)\sim 1/(pN)$.
In the intermediate regime, city~0 is in its phase of exponential growth,
and the mean arrival time can be estimated by imposing that
$\Lambda(\langle t_1\rangle)$ is of order unity.
We thus have
\beq
\langle t_1\rangle \approx
\frac{1}{\lambda-\mu}\,\ln\frac{\lambda-\mu}{p I_0}.
\label{tgrowth}
\eeq
The crossovers between these three regimes occur at
$p_1\approx(\lambda-\mu)/I_0$ and $p_2\approx(\lambda-\mu)/N$.
Figure~\ref{fig3} summarizes the above discussion,
whereas actual data will be shown in Figure~\ref{fig5}.

We now turn to the distribution of the outbreak time~$t^*_1$,
defined as the time of the arrival in city~1
of the first infected individual who induces an outbreak in city~1.
If the travel rate $p$ is small enough,
events where two or more infected individuals
coming from city~0 are simultaneously present in city~1 can be neglected.
Each infected individual entering city~1 has therefore a chance $1/\R_0$
(see~(\ref{pextinct})) to disappear before it induces an outbreak.
On the other hand,
the distribution of the arrival time $t_k$ of the $k$-th infected individual reads
\beq
\prob(t_k>t)=\P(N(t)<k)=\e^{-\Lambda(t)}\sum_{n=0}^{k-1}\frac{\Lambda(t)^n}{n!}.
\eeq
The probability that no outbreak occurred up to time $t$ is thus given by
\begin{eqnarray}
\avant \prob(t^*_1>t)\petit&=&\petit\sum_{k\ge1}(1-1/\R_0)\R_0^{1-k}\prob(t_k>t)\nonumber\\
\petit&=&\petit\e^{-(1-1/\R_0)\Lambda(t)}.
\end{eqnarray}
Comparing this expression with the corresponding one
for the first arrival~(\ref{eq:arrivaltime})
reveals that taking into account extinction
amounts to renormalizing $p$ to an effective rate $p^*$,
and accordingly $\Lambda(t)$ to $\Lambda^*(t)$,
multiplying both quantities by the scaling factor $1-1/\R_0$:
\beq
p^*=\left(1-\frac{1}{\R_0}\right)p,\quad
\Lambda^*(t)=\left(1-\frac{1}{\R_0}\right)\Lambda(t).
\label{renor}
\eeq
Within this approach, the distribution of the outbreak time in city~1 reads
\beq
\prob(t^*_1>t)=\e^{-\Lambda^*(t)}.
\label{tout}
\eeq

\subsection{Ballistic propagation on a one-dimensional array}

We now consider the stochastic SIS model in the one-dimensional geometry
of an infinite array of cities with initial populations $N$.
We still choose a symmetric travel rate $p$,
hence, in the course of time, these cities remain equally populated on average.

\begin{figure}[ht!]
\centering
\begin{tabular}{c}
\epsfig{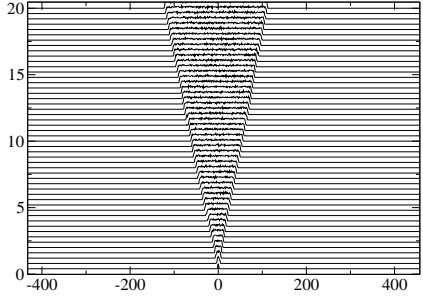}
\end{tabular}
\caption{ Propagation of a ballistic front in the SIS case ($\R_0=3$,
$p=0.01$, $N=100$) shown in the usual space-time representation
where the $y$-axis represents the time (in arbitrary units) and
where the $x$-axis represents the index of cities.}
\label{fig4}
\end{figure}

The model is observed to reach very quickly a ballistic propagation regime,
where the epidemic invades the whole array by propagating a ballistic front
moving at a well-defined finite velocity $V$.
Such a front is illustrated by the space-time plot of Figure \ref{fig4}.

The velocity $V$ of the front can be estimated as follows.
The time $T_n$ at which the epidemic reaches city $n$ can be recast as
\begin{equation}
T_n=\sum_{i=1}^nt^*_1(i),
\end{equation}
where $t^*_1(i)$ is the outbreak time in city $i$,
i.e., the arrival time of the infected individual which will trigger
the outbreak in that city,
with the origin of times being set to the
outbreak time in the previous city $i-1$.
Modeling the propagation from city $i-1$ to city $i$ by the case of two cities
studied above, we thus predict that the inverse velocity $1/V=\lim(T_n/n)$ reads
\begin{equation}
\frac{1}{V}\approx \langle t^*_1\rangle,
\label{vbalt}
\end{equation}
where $\langle t^*_1\rangle$ is the mean outbreak time of the problem of two cities,
with the natural initial condition $I_0=1$.

\begin{figure}[ht!]
\centering
\begin{tabular}{c}
\epsfig{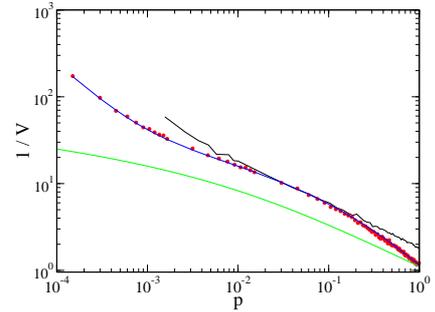}
\end{tabular}
\caption{
(Color online).
Upper black curve: Reciprocal of the ballistic velocity $V$ against $p$
($\R_0=3$, $N=100$, $p=0.01$).
Blue curve with red symbols: $\mean{t_1}$ of the two-city problem
(analytic prediction (\ref{t1_cox})) and numerical simulation, with $I_0=1$ and $p$ renormalized.
Lower smooth green curve: prediction (\ref{vsig}), (\ref{psig}) of FKPP theory.}
\label{fig5}
\end{figure}

In Figure~\ref{fig5} we show a comparison between
data for the reciprocal of the ballistic velocity $V$
and for $\langle t^*_1\rangle$, the mean outbreak time for two cities, with $I_0=1$.
The latter was measured in a Monte Carlo simulation as $\langle t_1\rangle$
with $p$ renormalized to $p^*$ (red symbols),
and computed using the analytic prediction (\ref{t1_cox}),
where $I_0(t)$ has the deterministic expression~(\ref{eq:determ}),
again with $p$ renormalized (blue curve).
The agreement validates the approximation (\ref{vbalt}).
The prediction (\ref{vsig}), (\ref{psig}) of the discrete FKPP theory is also shown.

It is indeed worth confronting the above analysis with yet another approach,
based on the analysis of the following deterministic equation for the densities
of infected individuals $\rho_i(t)= I_i(t)/N$:
\begin{equation}\label{fkppDisc}
\frac{\d \rho_i(t)}{\d t}=\lambda \rho_i(1-\rho_i)-\mu
\rho_i+p(\rho_{i+1}+\rho_{i-1}-2 \rho_i).
\end{equation}
This equation is the discrete version of the
FKPP equation~\cite{Fisher:1937,Kolmo:1937}.
Looking for a traveling-wave solution of the form $\rho_i(t) = f(i-Vt)$,
propagating with velocity $V$,
it follows that the function $f(x)$ obeys the differential-difference equation
\beqa
-V f'(x)\petit&=&\petit\lambda f(x)(1-f(x))-\mu f(x)\nonumber\\
\petit&+&\petit p\left (f(x+1)+f(x-1)-2f(x)\right).
\eeqa
If one assumes that the solution decays exponentially
as $f(x)\sim\e^{-\sigma x}$ for $x\to\infty$,
the velocity is found to depend continuously on $\sigma$, according to
\begin{equation}
V=\frac{\lambda-\mu+2p(\cosh\sigma-1)}{\sigma}.
\end{equation}
For a localized initial condition,
the actual velocity of the front is known~\cite{bramson,derridaspohn}
to be obtained by minimizing the above expression with respect to the spatial
decay rate $\sigma$.
This velocity reads
\begin{equation}
V=\frac{(\lambda-\mu)\sinh\sigma}{\sigma\sinh\sigma+1-\cosh\sigma},
\label{vsig}
\end{equation}
and it is reached for
\begin{equation}
p=\frac{\lambda-\mu}{2(\sigma\sinh\sigma+1-\cosh\sigma)}.
\label{psig}
\end{equation}
Both above equations give a parametric expression for the ballistic velocity $V$.
The usual prediction of the continuum FKPP equation,
i.e., $V=2((\lambda-\mu)p)^{1/2}$, is recovered as $\sigma\ll1$, i.e., $p\gg\lambda-\mu$.
The regime of most interest in the present context is the opposite one
($\sigma\gg1$, i.e., $p\ll\lambda-\mu$), where we have
\beq
\frac{1}{V}\approx\frac{1}{\lambda-\mu}\,\ln\frac{V}{p\e}.
\eeq
This regime is in correspondence with the estimate (\ref{tgrowth}) which holds in the intermediate growth regime.
Figure~\ref{fig5} shows that the discrete FKPP theory provides a good,
albeit not quantitative, description of the overall behavior of the ballistic velocity.

To conclude, in the one-dimensional SIS model, the epidemic always spreads ballistically,
with a non-zero velocity $V$. We expect this result to hold on any extended structure,
since it is merely a consequence of the existence of a very long-lived saturated regime
where a finite fraction of individuals are infected.
In a metapopulation model, this will for sure trigger an infection which will spread
over the whole network as long as $\R_0>1$.

\section{Metapopulation model in the SIR case}

We now turn to the case of the SIR model for the growth of an epidemic in a given city.
In contrast with the SIS model,
the number of infected individuals in a given city now falls off to zero at large times.
As a consequence, if the traveling rate is too small compared to the typical inverse
duration of the epidemic phase, the epidemic may die out in a city
before propagating to the neighboring ones.
This phenomenon will lead to the existence of a non-trivial threshold $p_\th$
for the travel rate, below which the disease will not spread in the system.
We will show that the problem can be recast in
terms of a bond percolation problem.
This picture greatly simplifies the analysis,
and leads to an estimate (lower bound) of the pandemic threshold.

\subsection{Stochastic SIR model for one city}

We first investigate the stochastic SIR process in the case of one
isolated city of population $N$. As in the SIS case, the process
is described at the population level (i.e., the individuals are indistinguishable),
but we now have three variables $S(t)$, $I(t)$, and $R(t)$
(number of recovered individuals), with $S(t)+I(t)+R(t)=N$.
The stochastic process is described by the reactions
\beq\label{sir1}
(S,I,R)\rightarrow\begin{cases}
(S-1,I+1,R)&\text{with rate}\ \lambda SI/N,\\
(S,I-1,R+1)&\text{with rate}\ \mu I,
\end{cases}
\eeq
with initial condition $I(0)=I_0$ and $R(0)=0$.
The second reaction corresponds to the recovery of an infected individual,
that stays immune later on.
The master equations for the process can be written in terms of two
independent integer random variables, say $S$ and $I$, and integrated numerically.

\begin{figure}[ht!]
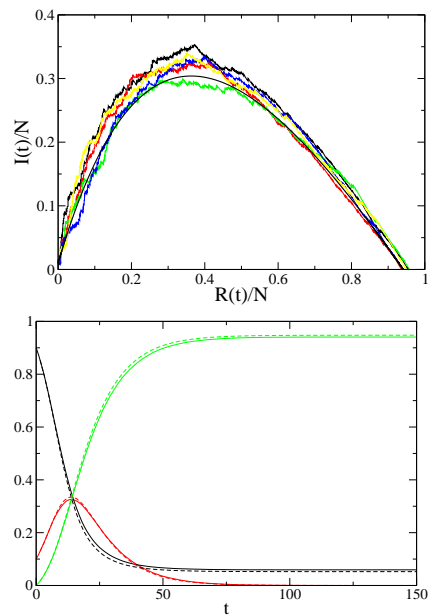

\centering
\begin{tabular}{c}
\epsfig{file=figure6a.eps,width=0.7\linewidth,clip=}\\
\epsfig{file=figure6b.eps,width=0.7\linewidth,clip=}
\end{tabular}
\caption{
(Color online).
SIR model for one city.
Top: five different realizations of the process plotted in the $I$-$R$ plane.
Smooth black line: solution of the deterministic equations (\ref{sirdyn}) ($N=1000$).
Bottom: densities of susceptible (black),infected (red), and removed (green) individuals
as a function of time.
Numerically measured mean densities (full) and corresponding
deterministic solutions (dashed) are hardly distinguishable ($\R_0=3$, $N=100$, $I_0=10$).}
\label{fig6}
\end{figure}

The deterministic equations describing the SIR process read
\beqa
\avant \frac{\d S}{\d t}\petit&=&\petit-\lambda\frac{SI}{N},\nonumber\\
\frac{\d I}{\d t}\petit&=&\petit\lambda\frac{SI}{N}-\mu I,\nonumber\\
\frac{\d R}{\d t}\petit&=&\petit\mu I.
\label{sirdyn}
\eeqa
At variance with the case of SIS, the above equations cannot be solved in closed form.
Figure~\ref{fig6} shows a comparison between the numerical solution of these equations
and a numerical simulation of the stochastic process.
The upper panel shows five different trajectories in the $I$-$R$ plane.
We observe relatively important fluctuations around the deterministic solution.
In particular the epidemic stops after a finite random time.
The lower panel demonstrates that mean quantities are however well described
by the deterministic approach.

A quantity of central interest for the sequel is the final size of the epidemic,
defined as the number $R(\infty)$ of recovered individuals after the epidemic has stopped:
\begin{equation}
R(\infty)=\mu\int_0^\infty\d\tau\,I(\tau).
\end{equation}
We refer the reader to~\cite{Bailey:1953,Ball:1994,Loef:1998} for
studies on the distribution of this quantity.
For a city with a large population ($N\gg1$),
$R(\infty)$ can be approximately determined in the framework of the deterministic approach.
It follows from~(\ref{sirdyn}) that
\begin{equation}
\frac{\d S}{\d R}=-\R_0\,\frac{S}{N},
\end{equation}
from which we obtain
\begin{equation}
\ln\frac{S(\infty)}{S(0)}=-\R_0\,\frac{R(\infty)}{N}.
\end{equation}
Using the initial condition given above, and the final condition $I(\infty)=0$, i.e.,
$S(\infty)+R(\infty)=N$,
we obtain a transcendental equation relating the densities $i_0=I_0/N$
and $r_\infty=R(\infty)/N$:
\begin{equation}
1-r_\infty=(1-i_0)\e^{-\R_0\,r_\infty},
\end{equation}
which reduces for $I(0)\ll N$, i.e., $i_0\ll1$, to
\begin{equation}
1-r_\infty=\e^{-\R_0 \,r_\infty}.
\end{equation}
The density $r_\infty$ starts rising linearly as
\beq
r_\infty\approx2(\R_0-1)\quad(\R_0\to1),
\label{rinfstart}
\eeq
and reaches unity exponentially fast for $\R_0\to\infty$, as $r_\infty\approx1-\e^{-\R_0}$.
For $\R_0=3$, the value used in numerical simulations, we have $r_\infty=0.940479$.

\subsection{Including travel: two cities}

\begin{figure}[ht!]
\centering
\begin{tabular}{c}
\epsfig{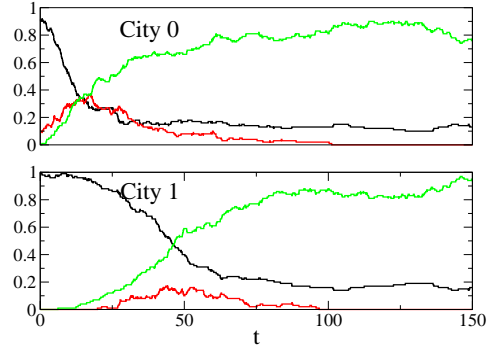}
\end{tabular}
\caption{
(Color online).
A typical history of the SIR process with travel between two cities:
plot of the densities of susceptible, infected, and removed individuals versus time
for cities~0 and 1.
Color code as in Figure~\ref{fig6}.}
\label{fig7}
\end{figure}

As in the SIS case, it is natural to start the study of propagation
by the case of two neighboring cities.
The SIR process inside each city is described in terms of three random
variables $S_i$, $I_i$, and $R_i$ ($i=0,1$), with $S_i+I_i+R_i=N_i$,
which evolve stochastically as in~(\ref{sir1}).
The traveling process between these two cities is given by the reactions
\beq\label{eq:travelsir}
(X_i,X_j)\rightarrow(X_i-1,X_j+1)\quad\text{with rate}\ \ p_{ij}X_i,
\eeq
where $X$ stands for $S$, $I$ or $R$.
We again assume symmetric diffusion.
A typical history of the system is shown in Figure~\ref{fig7}.

The first arrival time $t_1$ of an infected individual in city~1
is still distributed according to~(\ref{eq:arrivaltime}) and~(\ref{lambdadef}).
There is however a crucial difference between SIS and SIR.
As already discussed, in the present case of SIR,
the integrated rate $\Lambda(t)$ converges to a finite limit $\Lambda_\infty$
as $t\to\infty$.
For a large population $N$, the deterministic approach yields
\beq
\Lambda_\infty=\frac{N pr_\infty}{\mu}.
\eeq
As a consequence, there is now a non-zero probability $\exp(-\Lambda_\infty)$
that no infected individual travels from city~0 to city~1
during the whole epidemic in city~0.
In other words, the event that at least one infected individual reaches city~1,
i.e., that the time $t_1$ is finite,
only occurs with probability
\beq
\Pi=1-\e^{-\Lambda_\infty}=1-\e^{-N pr_\infty/\mu}.
\eeq
The distribution of $t_1$ is said to be defective~\cite{feller}.

Taking into account the fact that single infected individuals only trigger an outbreak
with probability $1-1/\R_0$,
the distribution of the outbreak time $t^*_1$ in city~1 is still given by~(\ref{tout}),
as a result of the renormalization procedure
which led us to replace $p$ by the effective rate $p^*$ (see~(\ref{renor})).
In particular, the probability of occurrence of an outbreak in city~1 reads
\beq
\Pi^*=1-\e^{-\Lambda^*_\infty}=1-\e^{-Np^*r_\infty/\mu}.
\label{eq:pi}
\eeq
This probability is non-trivial, i.e., less than one, in contrast with the SIS case.

\subsection{Propagation on extended structures}

We can now consider the general case where the cities are connected
so as to form a network or any other kind of extended structure
(one-dimensional array, regular finite-dimensional lattice, regular tree).
Individuals can travel by performing diffusion along the links of the network,
allowing thus the disease to spread over different cities.

The probability $\Pi^*$ given in~(\ref{eq:pi}) can be interpreted as the
probability that the disease propagates through one given link
from a city to one of its neighbors.
In the SIS case, since the integrated rate $\Lambda(t)$ diverges with time,
the probability $\Pi^*$ is trivially equal to unity.
In contrast, for the SIR process, this probability is less than unity, so that the disease can stop invading the network.

For a given network, denoting the bond percolation threshold by $p_c$,
the condition for the disease to spread is therefore
\begin{equation}
\Pi^*>p_c.
\label{eq:spread}
\end{equation}
This equation defines the {\it pandemic threshold}:
for $N$ large, (\ref{eq:spread}) yields $p>p_\th$, where
\beq
p_\th=\frac{\mu \abs{\ln(1-p_c)}}{N(1-1/\R_0)r_\infty}.
\label{pth}
\eeq

This static prediction is not claimed to give an exact value
for the pandemic threshold.
It can however be argued to provide a lower bound for the threshold,
which is also meant as a reasonable and useful estimate.
Indeed the picture of static percolation,
where links are occupied with constant probability $\Pi^*$,
corresponds to an ideal infinitely slow propagation,
where the epidemic can take an infinitely long time to cross some of the links.
In real situations, propagation is rather observed to take place with finite velocity.
This velocity can be thought of as providing a time cutoff $T$
for propagation across every single link of the network.
The effect of this cutoff time is to decrease the integrated rate from $\Lambda_\infty$
to the smaller value $\Lambda(T)$,
and hence to increase the threshold value of $p$
from the theoretical prediction~(\ref{pth}) to a higher value.

Let us now illustrate this result on some examples of networks
and other extended structures.

\subsubsection*{Cayley tree}

We consider first the geometry of a regular Cayley tree (or Bethe lattice)
with coordination number $k$, so that $p_c=1/(k-1)$.

\begin{figure}[ht!]
\centering
\begin{tabular}{c}
\epsfig{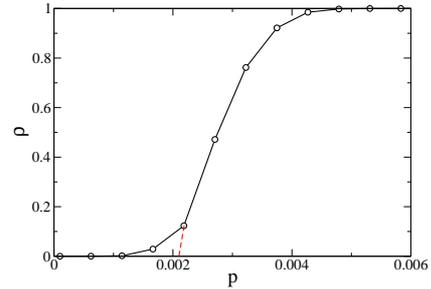}
\end{tabular}
\caption{
(Color online).
Fraction of infected cities at a very large time
versus travel probability $p$ on the Cayley tree
($k=3$, $t=5\times10^5$, $\R_0=3$, $N=100$).
The red dashed line points toward a threshold value near $p_\th\approx2.1\times10^{-3}$.}
\label{fig8}
\end{figure}

For the simplest example of $k=3$,
and an epidemic initially located at the center of a tree with 15 generations,
Figure~\ref{fig8} shows a plot of the fraction of infected cities at a very long time.
We indeed observe a pandemic threshold near $p_\th\approx2.1\times10^{-3}$.
For $p_c=1/2$, $\mu=0.1$, $\R_0=3$, hence $r_\infty=0.940479$, and $N=100$,
the predicted static threshold is $p_\th=1.105\times10^{-3}$.
The observed threshold is thus some two times larger than the static one.

\subsubsection*{Square lattice}

We now consider the two-dimensional situation of propagation on the square lattice.
This lattice also has bond percolation threshold $p_c=1/2$,
so that the above static threshold of $p_\th=1.105\times10^{-3}$
still holds for the same parameter values.

\begin{figure}[ht!]
\centering
\begin{tabular}{c}
\epsfig{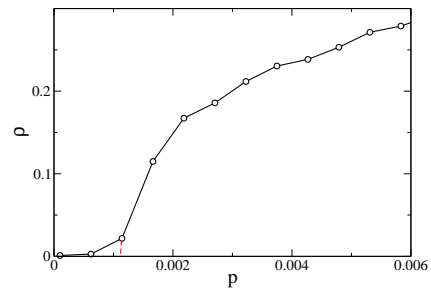}
\end{tabular}
\caption{
(Color online).
Fraction of infected cities at a very large time
versus travel probability $p$ on the square lattice.
Same parameters as in Figure~\ref{fig8}.
The red dashed line points toward a threshold value near $p_\th\approx1.1\times10^{-3}$.}
\label{fig9}
\end{figure}

\begin{figure}[ht!]
\centering
\begin{tabular}{c}
\epsfig{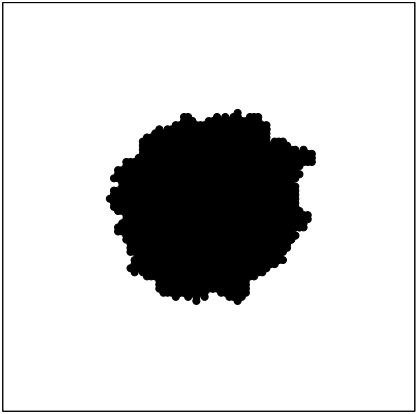}
\end{tabular}
\caption{
Typical infected region on the square lattice after a large but finite time
(system size $50\times 50$, $N=100$, $p=0.01\gg p_\th$).}
\label{fig10}
\end{figure}

On a $50\times50$ array,
the data of Figure~\ref{fig9} clearly demonstrates the existence of a threshold
near $p_\th\approx1.1\times10^{-3}$,
in good quantitative agreement (within ten percent, say) with the predicted static value.
In figure \ref{fig10} we provide an example of the observed shape of the infected region
for a large but finite time.

\subsubsection*{One-dimensional array}

We finally address the case of propagation along an infinite array of cities.
The percolation threshold in the one-dimensional case is $p_c=1$.
Our analysis therefore suggests that the disease will always stop,
after having invaded only a finite range of typical size $\xi$, but not the whole array.
The static percolation approach suggests that $\xi$ diverges as $\Pi^*\to1$
in the same way as the static correlation length in percolation theory,
namely $\xi\sim1/(1-\Pi^*)$.
We thus find the exponential growth
\begin{equation}
\xi\sim\exp\left(\frac{pN(1-1/\R_0)r_\infty}{\mu}\right).
\end{equation}

\begin{figure}[ht!]
\centering
\begin{tabular}{c}
\epsfig{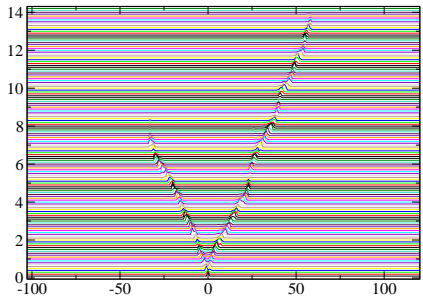}
\end{tabular}
\caption{
(Color online).
Propagation of a ballistic front in the SIR case ($\R_0=3$, $p=0.01$, $N=100$).
Time (arbitrary units) versus index of the city.
The front is observed to stop at a random time.}
\label{fig11}
\end{figure}

\begin{figure}[ht!]
\centering
\begin{tabular}{c}
\epsfig{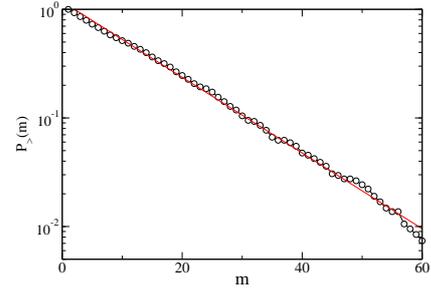}
\end{tabular}
\caption{
(Color online).
Cumulative distribution of the location $m$ of the rightmost infected city ($N=100$).
The slope of the fitted red line yields $\xi=12.4$.}
\label{fig12}
\end{figure}

\begin{figure}[ht!]
\centering
\begin{tabular}{c}
\epsfig{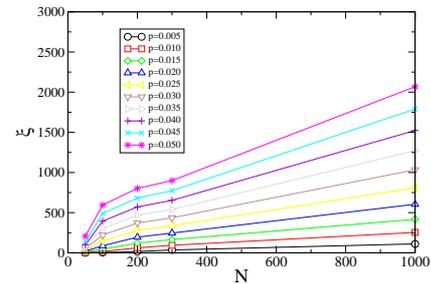}
\end{tabular}
\caption{
(Color online).
Characteristic length $\xi$ versus population $N$ of one city.
Colors correspond to various values of $p$.}
\label{fig13}
\end{figure}

We indeed observe a symmetric ballistic front on both sides of the seed.
Each branch suddenly dies at a random time,
and hence in a randomly located city, say number $m$.
Such a front is shown in Figure~\ref{fig11}.
For given parameter values,
the right stopping point $m$ is observed to be exponentially distributed,
with a cumulative distribution falling off as $\exp(-m/\xi)$ (see Figure~\ref{fig12}).
This exponential distribution confirms our intuition
that a local mechanism is responsible for the arrest of the propagation.
The $p$ and $N$ dependence of the characteristic length $\xi$ thus measured is shown in Figure~\ref{fig13}.
The data demonstrate an increase of $\xi$ with both $p$ and $N$.
The agreement with the static prediction however remains very qualitative.

\subsubsection*{Scale-Free networks}

We close up our analysis of the propagation on a SIR epidemic
by considering the case of a scale-free network.
For an uncorrelated network, the percolation threshold is
known to be given by $p_c=\langle k\rangle/(\langle k^2\rangle-\langle k\rangle)$
(see~(\ref{pcnet})).
As most scale-free networks are such that $\mean{k^2}\gg\mean{k}$,
we have $p_c\approx\langle k\rangle/\langle k^2\rangle\ll1$,
so that our static prediction~(\ref{pth}) for the pandemic threshold reads
\begin{equation}
p_\th\approx\frac{\mu}{N}
\frac{\langle k\rangle}{\langle k^2\rangle}
\frac{1}{(1-1/\R_0)r_\infty}.
\label{pth1}
\end{equation}
In the same regime, it is worth recasting the result
of~\cite{Colizza:2007a}, where the expression for $\R_*$ is recalled
in~(\ref{eq:rstar}), in the form of a threshold value $\hat p_\th$ of
the travel rate. Using the fact that $\alpha=r_\infty$ for SIR, we
obtain
\begin{equation} p_\th=\frac{\R_0}{\mean{k}}\,\hat p_\th.
\end{equation}
Our result thus only differs from that of~\cite{Colizza:2007a} by an
inessential multiplicative factor involving the mean degree $\mean{k}$
and the reproductive rate $\R_0$. The coincidence is all the more
striking that both approaches are largely different. Both expressions
agree to predict that the threshold diverges quadratically as
$1/(\R_0-1)^2$ as the basic reproductive rate $\R_0\to1$. Indeed
$r_\infty$ vanishes linearly (see~(\ref{rinfstart})).

\section{Conclusion}

In this paper we have considered the metapopulation models for the SIS
and SIR processes in their stochastic versions. We have put forward a
description considering individuals as indistinguishable, but keeping
the full stochastic character of the populations in the various
compartments. This intermediate level of description, along the lines
of the theory of urn models and migration processes, allows very
efficient numerical simulations. We have considered with great care
temporal effects in the spread of a disease from one city to another
and investigated its propagation at a global scale on scale-free
networks and other extended structures such as regular trees and lattices.

For the SIS case we always observe a ballistic
propagation. For the SIR case, even after an infinitely long time,
the disease only gets transmitted with probability $\Pi^*<1$ through
every link of the network. This picture allows us to rephrase the
possibility of global invasion as a static bond percolation problem.
The resulting prediction is an estimate (lower bound) for the pandemic
threshold, expressed as a threshold value $p_\th$ for the travel rate $p$. In
the case of scale-free networks, our prediction is virtually identical
to that of~\cite{Colizza:2007a}.
Our approach is however not limited to
the case $\R_0\rightarrow 1$, and takes into account both temporal and
topological fluctuations. For other geometries (Cayley tree, square
lattice), our static prediction yields a reasonable estimate for the
pandemic thresholds observed in numerical simulations.

As in most previous theoretical studies, we assumed that the
populations of all cities were equal and that the traveling rate per
link and per individual was constant. It would be interesting to
extend our result and to test for the relevance of travel and/or
population heterogeneities on the existence of a pandemic threshold.

Finally, it is worth putting the present work in a broader perspective.
The relationship between epidemic spreading and percolation has already
been discussed in the context of contact networks.
Table~\ref{comparaison} summarizes the parallel between
the contact network and the metapopulation approaches.
It has been emphasized first by Grassberger that the spreading of an epidemic
over a contact network can be mapped onto directed percolation~\cite{Grass:1983}.
It was then realized that, as a general rule, epidemic models without immunization
belong to the universality class of directed percolation,
whereas those with immunization, such as SIR,
belong to the universality class of dynamical percolation
(see~\cite{hinrichsen} for a review).
In the latter case, clusters of immune individuals
have the same critical properties as usual percolation clusters.
The effective description of epidemic spreading put forward in this work
demonstrates that epidemic spreading is also intimately related
to static percolation at the metapopulation level.

\begin{table}[!ht]
\begin{center}
\begin{tabular}{|c|c|c|}
\hline
Network&Nodes&Links\\
\hline
contact&individuals&contacts\\
\hline
metapopulation&cities&travel\\
\hline
\end{tabular}
\end{center}
\caption{A comparison between models of epidemic spreading
defined on contact networks and at the metapopulation level.}
\label{comparaison}
\end{table}

\appendix
\section{Pandemic threshold on a network with arbitrary travel rates}

In this appendix we determine the pandemic threshold $p_\th$
on an uncorrelated network where travel is described by arbitrary hopping rates.

We consider an uncorrelated network,
denote by $P_k$ be the degree distribution of the nodes,
and use the framework of degree classes,
assuming that all nodes with given degree $k$ are equivalent.
Travel is defined by the rate $p_{kl}$
for an individual to hop from a node with degree $k$ to a neighboring node of degree $l$.
The stationary populations $N_k$ obey the balance equation
\beq
N_k\sum_l\w P_l p_{kl}=\sum_l \w P_l p_{lk} N_l,
\eeq
where
\beq
\w P_k=\frac{kP_k}{\mean{k}}
\eeq
is the degree distribution of a neighboring node of a given node.
For instance, uniform rates $p_{kl}=p$ yield uniform populations $N_k=N$,
whereas the rates $p_{kl}=p/k$ of ordinary random walk yield populations $N_k=kN$.
More generally, separable (i.e., factorized) travel rates of the form $p_{kl}=a_kb_l$
yield stationary populations $N_k=N\,b_k/a_k$.

The basic quantity in the metapopulation model is the probability
that the epidemic will propagate (in an infinitely long time)
from a node of degree $k$ to a node of degree $l$.
Dropping the star for conciseness, this probability reads
\beq
\Pi_{kl}=1-\exp\left(-\frac{N_kp_{kl}(1-1/\R_0)r_\infty}{\mu}\right).
\eeq
In order to determine the static pandemic threshold of SIR on the network,
we are thus led to consider the problem of directed bond percolation on the network
defined by the above probabilities $\Pi_{kl}$ that an oriented link is open
from a node of degree $k$ to a node of degree~$l$.
Let us introduce the probabilities $Q_{kl}$ that an oriented link
from a node of degree $k$ to a node of degree $l$
leads to a node belonging to the giant component.
These quantities obey
\beq
Q_{kl}=\Pi_{kl}\left(1-\prod_v(1-Q_{l,m(v)})\right),
\label{qeq}
\eeq
where $v$ runs over the $(l-1)$ neighbors of the node of degree $l$
which are not the initial one and $m(v)$ are their degrees.
Near the threshold, the probabilities $Q_{kl}$ are expected to be small.
Linearizing the above relation, we obtain
\beq
Q_{kl}=(l-1)\Pi_{kl}\sum_m\w P_m Q_{lm}.
\eeq
Setting $Q_{kl}=\Pi_{kl}G_l/\w P_l$,
the above condition reads $G_l=\sum_mM_{lm}G_m$, with
\beq
M_{kl}=\frac{k(k-1)}{\mean{k}}\,P_k\,\Pi_{kl}.
\eeq
The percolation threshold
is therefore given by the condition that the largest (Perron-Frobenius) eigenvalue
of the positive matrix $M$ equals unity.

The usual percolation problem
corresponds to the case where every link is occupied with probability $p$,
i.e., $\Pi_{kl}=p$ for all $k$ and $l$.
We thus recover the known result
\beq
p_c=\frac{\mean{k}}{\mean{k(k-1}}.
\label{pcnet}
\eeq
In particular, for a regular Cayley tree with coordination number $k$, we have
\beq
p_c=\frac{1}{k-1}.
\eeq
The reasoning leading to~(\ref{qeq}) can be extended to other cases.
Let us mention the example of a bipartite tree,
where nodes with degree $k_1$ are neighbors of nodes of degree $k_2$.
We then obtain
\beq
p_c=\frac{1}{\sqrt{(k_1-1)(k_2-1)}}.
\eeq
For separable probabilities, of the form $\Pi_{kl}=q_kr_l$,
we obtain the following threshold condition
\beq
\frac{\mean{k(k-1)q_kr_k}}{\mean{k}}=1.
\eeq
In a more general case, we have to resort to a numerical computation
of the largest eigenvalue of the matrix~$M$.


\begin{thebibliography}{99}

\bibitem{Rvachev:1985} L.A. Rvachev and I.M. Longini, A mathematical model for the global spread of Influenza, {\it Math. Biosci.} {\bf 75}, 3-23 (1985).

\bibitem{Longini:1988} I.M. Longini, A mathematical model for predicting the geographic spread of new infectious agents, {\it Math. Biosci.} {\bf 90}, 367-383 (1988).

\bibitem{Grais:2003} R.F. Grais, J. Hugh Ellis, and G.E. Glass, Assessing the impact of airline travel on the geographic spread of pandemic influenza, {\it Eur. J. Epidemiol.} {\bf 18}, 1065-1072 (2003).

\bibitem{Grais:2004} R.F. Grais, J. Hugh Ellis, A. Kress, and G.E. Glass, Modeling the spread of annual influenza epidemics in the US: the potential role of air travel {\it Health Care Manage. Sci.} {\bf 7}, 127-134 (2004).

\bibitem{Colizza:2006a} V. Colizza, A. Barrat, M. Barth\'elemy, and A. Vespignani, The Modeling of Global Epidemics: Stochastic Dynamics and Predictability, {\it Bull. Math. Biol.} {\bf 68}, 1893-1921 (2006).

\bibitem{Hufnagel:2004} L. Hufnagel, D. Brockmann, and T. Geisel, Forecast and control of epidemics in a globalized world, {\it Proc. Natl Acad. Sci. (USA)} {\bf 101}, 15124-15129 (2004).

\bibitem{Colizza:2007} V. Colizza, A. Barrat, M. Barth\'elemy, and A. Vespignani, Predictability and epidemic pathways in global outbreaks of infectious diseases: the SARS case study, {\it BMC Medicine} {\bf 5}, 34 (2007).

\bibitem{Flahault:1991} A. Flahault and A.-J. Valleron, A method for assessing the global spread of HIV-1 infection based on air-travel, {\it Math. Pop. Studies} {\bf 3}, 1-11 (1991).

\bibitem{Vespi:2009a} P. Bajardi, C. Poletto, D. Balcan, H. Hu, B. Goncalves, J.J. Ramasco, D. Paolotti, N. Perra, M. Tizzoni, W. Van den Broeck, V. Colizza, and A. Vespignani, Modeling vaccination campaigns and the Fall/Winter $2009$ activity of the new A(H1N1) influenza in the Northern Hemisphere, {\it Emerging Health Threats Journal} {\bf 2}, 11 (2009).

\bibitem{Vespi:2009b} D. Balcan, H. Hu, B. Goncalves, P. Bajardi, C. Poletto, J.J. Ramasco, D. Paolotti, N. Perra, M. Tizzoni, W. Van den Broeck, V. Colizza, and A. Vespignani, Seasonal transmission potential and activity peaks of the new influenza A(H1N1): a Monte Carlo likelihood analysis based on human mobility, {\it BMC Medicine} {\bf 7}, 45 (2009).

\bibitem{Colizza:2006b} V. Colizza, A. Barrat, M. Barth\'elemy, and A. Vespignani, The role of the airline transportation network in the prediction and predictability of global epidemics, {\it Proc. Natl Acad. Sci. (USA)} {\bf 103}, 2015-2020 (2006).

\bibitem{Colizza:2007a} V. Colizza and A. Vespignani, Invasion threshold in heterogeneous metapopulation networks, {\it Phys. Rev. Lett.} {\bf 99}, 148701 (2007).

\bibitem{Colizza:2007b} V. Colizza and R. Pastor-Satorras, Reaction-diffusion processes and metapopulation models in heterogeneous networks, {\it Nature Phys.} {\bf 3}, 276-282 (2007).

\bibitem{Colizza:2008} V. Colizza and A. Vespignani, Epidemic modeling in metapopulation systems with heterogeneous coupling pattern: Theory and simulations, {\it J. Theor. Biol.} {\bf 251}, 450-467 (2008).

\bibitem{Gardiner:2004} W.C. Gardiner, Handbook of Stochastic Methods for Physics, Chemistry, and Natural Sciences, 3rd edition (Springer, New York, 2004).

\bibitem{Godreche:2002} C. Godr\`eche and J.M. Luck, Nonequilibrium dynamics of urn models, {\it J. Phys.: Condens. Matter} {\bf 14}, 1601-1615 (2002).

\bibitem{Godreche:2007} C. Godr\`eche, From Urn Models to Zero-Range Processes: Statics and Dynamics, in {\it Ageing and the Glass Transition} (Lecture Notes in Physics {\bf 716}) (Springer, Berlin, 2007), p. 261.

\bibitem{Allen:2003} L.J.S. Allen, An Introduction to Stochastic Processes with Applications to Biology (Pearson, Prentice Hall, New Jersey, 2003).

\bibitem{Fisher:1937} R.A. Fisher, The wave of advance of advantageous genes, {\it Ann. Eugenics} {\bf 7}, 353-369 (1937).

\bibitem{Kolmo:1937} A. Kolmogorov, I. Petrovsky, and N. Piscounov, Etude de l'\'equation de la diffusion avec croissance de la quantit\'e de mati\`ere et son application \`a un probl\`eme biologique, {\it Moscow Univ. Bull. Math.} {\bf 1}, 1 (1937).

\bibitem{feller} W. Feller, An introduction to probability theory and its applications (Wiley and Sons, New York, 1968).

\bibitem{karlin} S. Karlin and H.M. Taylor, A First Course in Stochastic Processes (Academic Press, 1975).

\bibitem{kendall} D.G. Kendall, On the generalized birth-and-death process, {\it Annals of Math. Stat.} {\bf 19}, 1-15 (1948).

\bibitem{Bailey} N.T.J. Bailey, The Mathematical Theory of Infectious Diseases and its Applications, 2nd ed. (Griffin, London, 1975).

\bibitem{Gautreau:2008} A. Gautreau, A. Barrat, and M. Barth\'elemy, Global disease spread: statistics and estimation of arrival times, {\it J. Theor. Biol.} {\bf 251}, 509-522 (2008).

\bibitem{Cox} D.R. Cox, Some statistical methods connected with series of events, {\it J. R. Statist. Soc. Ser. B} {\bf 17}, 129-164 (1955).

\bibitem{bramson} M.D. Bramson, Convergence of solutions of the Kolmogorov equation to travelling waves, {\it Mem. Amer. Math. Soc.} {\bf 285}, 1-190 (1983).

\bibitem{derridaspohn} B. Derrida and H. Spohn, Polymers on Disordered Trees, Spin Glasses, and Traveling Waves, {\it J. Stat. Phys.} {\bf 51}, 817-840 (1988).

\bibitem{Bailey:1953} N.T.J. Bailey, The total size of a general stochastic epidemic, {\it Biometrika} {\bf 40}, 177-185 (1953).

\bibitem{Ball:1994} F. Ball and I. Nasell, The shape of the size distribution of an epidemic in a finite population, {\it Math. Biosci.} {\bf 123}, 167-181 (1994).

\bibitem{Loef:1998} A. Martin-Loef, The final size of a nearly critical epidemic, and the first passage time of a Wiener process to a parabolic barrier, {\it J. Appl. Prob.} {\bf 35}, 671-682 (1998).

\bibitem{Grass:1983} P. Grassberger, On the critical behavior of the general epidemic process and dynamical percolation, {\it Math. Biosc.} {\bf 63}, 157-172 (1983).

\bibitem{hinrichsen} H. Hinrichsen, Non-equilibrium critical phenomena and phase transitions into absorbing states, {\it Adv. Phys.} {\bf 49}, 815-958 (2000).


\end{thebibliography}
\end{document}